\pdfoutput=1

\documentclass[11pt]{article}
\usepackage[preprint]{acl}
\usepackage{amsmath}
\usepackage{booktabs}
\usepackage[normalem]{ulem}
\useunder{\uline}{\ul}{}
\usepackage{times}
\usepackage{latexsym}

\usepackage[T1]{fontenc}

\usepackage[utf8]{inputenc}

\usepackage{microtype}

\usepackage{inconsolata}

\usepackage{graphicx}
\usepackage{adjustbox}
\usepackage{multirow}
\newenvironment{sizeddisplay}[1]
 {\par\nopagebreak#1\noindent\ignorespaces}
 {\nopagebreak\ignorespacesafterend}
\title{Optimizing Query Generation for Enhanced Document Retrieval in RAG}

\author{Hamin Koo\thanks{This work was done while the author was an intern at KAIST MLAI.} \\
  Independent
\\
  \texttt{hamin2065@google.com} \\\And
  Minseon Kim \\
  KAIST \\
  \texttt{minseonkim@kaist.ac.kr} \\\And
  Sung Ju Hwang \\
  KAIST, DeepAuto.ai \\
  \texttt{sjhwang82@kaist.ac.kr} \\}

\begin{document}

\maketitle

\begin{abstract}
Large Language Models (LLMs) excel in various language tasks but they often generate incorrect information, a phenomenon known as "hallucinations". Retrieval-Augmented Generation (RAG) aims to mitigate this by using document retrieval for accurate responses. However, RAG still faces hallucinations due to vague queries. This study aims to improve RAG by optimizing query generation with a query-document alignment score, refining queries using LLMs for better precision and efficiency of document retrieval. Experiments have shown that our approach improves document retrieval, resulting in an average accuracy gain of 1.6\%.
\end{abstract}
\section{Introduction}

Although Large Language Models (LLMs) demonstrate surprising performance in diverse language tasks, hallucinations in LLMs have become an increasingly critical problem. Hallucinations occur when LLMs generate incorrect or misleading information, which can significantly undermine their reliability and usefulness. One approach to mitigate this problem is Retrieval-Augmented Generation (RAG) \citep{lewis2021retrievalaugmentedgenerationknowledgeintensivenlp}, which leverages document retrieval to provide more accurate answers to user queries by grounding the generated responses in factual information from retrieved documents.

\begin{figure}[t]
\vspace{-0.1in}
\includegraphics[width=\linewidth]{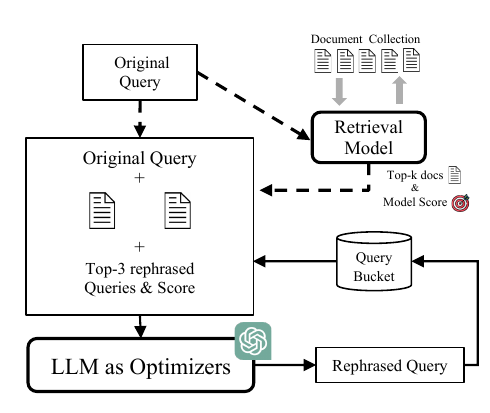}
\vspace{-0.35in}
\caption{\small Concept figure of QOQA. Given expansion query with top-k docs, we add top-3 rephrased queries and scores to LLM. We optimize the query based on the scores and generate the rephrased query.}
\label{figure:1}
\vspace{-0.25in}
\end{figure}
However, an incomplete RAG system often induces hallucinations due to vague queries that fail to accurately capture the user's intent~\citep{zhang2023siren}, highlighting a significant limitation of RAG in LLMs~\citep{niu2024ragtruthhallucinationcorpusdeveloping,wu2024clashevalquantifyingtugofwarllms}. The performance of RAG heavily depends on the clarity of the queries, with short or ambiguous queries negatively impacting search results ~\citep{jagerman2023queryexpansionpromptinglarge}. Recent studies~\citep{wang2023query2doc, jagerman2023queryexpansionpromptinglarge} have demonstrated that query expansion using LLMs can enhance the retrieval of relevant documents. Pseudo Relevance Feedback (PRF)~\citep{10.1145/383952.383972, 10.1145/1645953.1646259} further refines search results by automatically modifying the initial query based on top-ranked documents, without requiring explicit user input. By assuming the top results are relevant, PRF enhances the query, thereby improving the accuracy of subsequent retrievals.

To address this issue, our goal is to generate concrete and precise queries for document retrieval in RAG systems by optimizing the query. We propose \textbf{Q}uery \textbf{O}ptimization using \textbf{Q}uery exp\textbf{A}nsion (\textbf{QOQA}) for precise query for RAG systems. We employ a top-k averaged query-document alignment score to refine the query using LLMs. This approach is computationally efficient and improves the precision of document retrieval, thereby reducing hallucinations. In our experiments, we demonstrate that our approach enables the extraction of correct documents with an average gain of 1.6\%.
\vspace{-0.05in}
\section{Related Works}
\vspace{-0.05in}
\paragraph{Hallucination in RAG}
Despite the vast training data of large language models (LLMs), the issue of hallucination of LLM continues to undermine user belief. Among the strategies to mitigate, the Retrieval-Augmented Generation (RAG) method has proven effective in reducing hallucinations, enhancing the reliability and factual consistency of LLM outputs, thus ensuring accuracy and relevance in response to user queries~\citep{shuster2021retrievalaugmentationreduceshallucination, béchard2024reducinghallucinationstructuredoutputs}. However, RAG does not thoroughly eliminate hallucinations~\citep{béchard2024reducinghallucinationstructuredoutputs, niu2024ragtruthhallucinationcorpusdeveloping} that encouraged further refined RAG systems for lowered hallucination.
LLM-Augmenter~\citep{peng2023checkfactstryagain} leverages external knowledge and automated feedback via Plug and Play~\citep{li2024selfcheckerplugandplaymodulesfactchecking} modules to enhance model responses. Moreover, EVER~\citep{kang2024evermitigatinghallucinationlarge} introduces a real-time, step-wise generation and hallucination rectification strategy that validates each sentence during generation, preventing the propagation of errors.

\vspace{-0.2in}
\paragraph{Query Expansion}
Query expansion improves search results by modifying the original query with additional relevant terms, helping to connect the user's query with relevant documents. There are two primary query expansion approaches: retriever-based and generation-based. Retriever-based approaches expand queries by using results from a retriever, while generation-based methods use external data, such as large language models (LLMs), to enhance queries. 

Several works~\citep{wang2023query2doc, 10.1145/3539618.3591992,jagerman2023queryexpansionpromptinglarge} leverage LLMs for expanding queries. Query2Doc~\citep{wang2023query2doc} demonstrated that LLM-generated outputs added to a query significantly outperformed simple retrievers. However, this approach can introduce inaccuracies, misalignment with target documents, and highly susceptibility to LLM hallucinations. Retrieval-based methods~\cite{10.1145/1835449.1835546, 10.5555/1760167.1760198, li2023improvingqueryrepresentationsdense, lei2024corpus} enhance search query effectiveness by incorporating related terms or phrases, enriching the query with relevant information. Specifically, CSQE~\citep{lei2024corpus} uses an LLM to extract key sentences from retrieved documents for query expansion, creating task-adaptive queries, although this can lead to excessively long queries. When comparing CSQE-expanded queries with those evaluated by BM25~\citep{10.1561/1500000019} and re-ranked using a cross-encoder~\citep{wang2020minilmdeepselfattentiondistillation} from BEIR~\citep{thakur2021beir}, the performance improvement is minimal.
\section{Query Optimization using Query Expansion}

\subsection{Query optimization with LLM}
\label{sec:1}
To optimize the query, we utilize a Large Language Model (LLM) to rephrase the query based on its score. Initially, we input the original query and retrieve $N$ documents using a retriever. Next, we concatenate the original query with the top $N$ retrieved documents to create an expanded query, which is then sent to the LLM to generate $R_0$ rephrased queries. These rephrased queries are evaluated for alignment with the retrieved documents, and the pair of query-document alignment scores and queries are stored in a query bucket. The alignment score is determined using a retrieval model that measures the correlation between the query and the retrieved documents (Section~\ref{sec:2}).

\begin{figure}[t]
\vspace{-0.1in}
\centering
\includegraphics[width=0.85\linewidth]{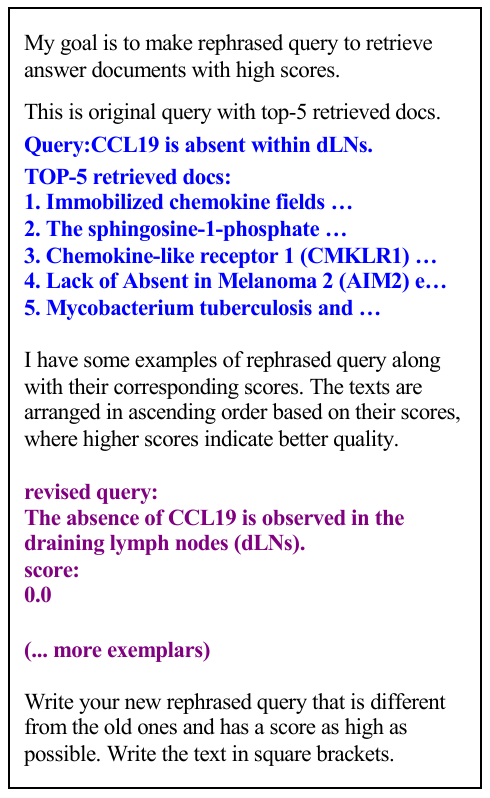}
\vspace{-0.15in}
\caption{\small Prompt template used in QOQA. The black texts describe instructions of the optimizing task. The {\color{blue}blue} texts are original query with top-$N$ retrieved documents with the original query. The {\color{purple}purple} texts are revised queries by LLM optimizer and scores.}
\label{figure:2}
\vspace{-0.25in}
\end{figure}

We update the prompt template with the original query, the retrieved documents, and the top $K$ rephrased queries, as illustrated in Figure~\ref{figure:2}. To ensure improved performance than original query, we always include the original query information in the template. In the later optimization steps $i$, based on the scores, we generate a $R_i$ rephrased query and add it to the query bucket.

\subsection{Query-document alignment score}
\label{sec:2}
To employ query-document alignment score in optimization step, we use three types of evaluation scores: BM25 scores from sparse retrievals, dense scores from dense retrievals, hybrid scores that combine the sparse and dense retrievals. 

Given query $q_i$, and documents set $D=\{d_j\}_{j=1}^J$ the BM25 alignment score is as follow,
\begin{sizeddisplay}{\footnotesize}
\centering
\begin{align}
\texttt{BM25}(q_i, D) = \frac{\texttt{IDF}(q_i) \cdot \texttt{f}(q_i, D) \cdot (k_1 + 1)}{\texttt{f}(q_i, D) + k_1 \cdot (1 - b + b \cdot \frac{|D|}{\textsc{avgDL}})} 
\end{align}
\end{sizeddisplay}
where $\texttt{f}(q_i, D)$ is frequency of query terms in the document $D$, $|D|$ is the length of the document, $\textsc{avgDL}$ is average document length, and $k_1$ and $b$ are default hyper-parameters from Pyserini~\citep{Lin_etal_SIGIR2021_Pyserini}. $\texttt{IDF}(q_i)$ is inverse document frequency term as follow,
\begin{equation}
\texttt{IDF}(q_i) = \log \frac{N - n(q_i) + 0.5}{n(q_i) + 0.5} 
\end{equation}
where $\texttt{IDF}(q_i)$ is calculated with total number of documents $N$, and $n(q_i)$ as number of documents containing $q_i$. 

Dense score is relevance score between queries and documents using learned dense representations, i.e., embedding space. As both queries and documents are embedded into the high-dimensional continuous vector space, alignment score $\texttt{Dense}$ is calculated as follow,
\begin{equation}
\texttt{Dense}(q_i, d_j) = E_{q_i} \cdot E_{d_j}
\end{equation}
where $E_{q_i}$ and $E_{d_j}$ are the dense embedding vectors of the query $q_i$ and the document $d_j \in D$, respectively, from dense retrieval model. For our experiment, we employ BAAI/bge-base-en-v1.5~\citep{xiao2024cpackpackagedresourcesadvance} model.

Hybrid score combines both $\texttt{BM25}$ scores and $\texttt{Dense}$ scores by appropriately tuning parameters of alpha $\alpha$ as follow,
\begin{equation}
\texttt{Hybrid}(q_i, d_j) = \alpha \cdot \texttt{BM25}(q_i, D) + \texttt{Dense}(q_i, d_j).
\end{equation}

\section{Results}

\paragraph{Dataset}
We evaluate on three retrieval datasets from BEIR~\citep{thakur2021beir}: SciFact~\citep{wadden-etal-2020-fact}, Trec-Covid~\citep{10.1145/3451964.3451965} and FiQA~\citep{10.1145/3184558.3192301}. We evaluated on fact checking task about scientific claims, Bio-medical information retrieval, and question answering task on financial domain, respectively. 

\begin{table}[t]
\caption{\small Results of document retrieval task. All scores denote nDCG@10. \textbf{Bold} indicates the best result across all models, and the second best is \underline{underlined}.}
\centering
\vspace{-0.1in}
\begin{adjustbox}{width=\linewidth}
  \begin{tabular}{lccc}
        \toprule
        & Scifact  & Trec-covid & FiQA  \\
        \midrule
        \textit{{Sparse Retrieval}}   &&&\\
        BM25                                  & 67.9                 & 59.5                 & 23.6                 \\
        + RM3~\citep{10.1145/1645953.1646259}                                  & 64.6                 & 59.3                 & 19.2                 \\
        + Q2D/PRF~\citep{jagerman2023queryexpansionpromptinglarge} & 71.7                 & 73.8                 & 29.0                 \\
        + CSQE~\citep{lei2024corpus}                                 & 69.6                 & 74.2                 & 25.0                 \\
        + QOQA (BM25 score)                       & 67.5                 & 61.1                 & 21.4                 \\
        + QOQA (Dense score)                      & 69.7                 & 48.4                 & 23.6                 \\
        + QOQA (Hybrid score) & 66.4                 & 43.2                 & 22.4                 \\ 
        \midrule
        \textit{{Dense Retrieval}}& &&\\
        BGE-base-1.5                          & 74.1                 & {\ul 78.2}           & \textbf{40.7}        \\
        + CSQE~\citep{lei2024corpus}                                 & 73.7                 & {\ul 78.2}           & 40.1                 \\
        + QOQA (BM25 score)                       & \textbf{75.4}        & 60.6                 & 37.4                 \\
        + QOQA (Dense score)                      & {\ul 74.3}           & 77.9                 & {\ul 40.6}           \\
        + QOQA (Hybrid score)                     & 73.9                 & \textbf{79.2}        & 40.0                \\ 
        \bottomrule
    \end{tabular}
    \label{table:1}
\end{adjustbox}
\vspace{-0.15in}
\end{table}

\begin{table*}[t]
\vspace{-0.1in}
\caption{\small Examples from SciFact, and FiQA dataset. {\color{blue}Blue} texts are overlapping keywords between answer document and rephrased query.}
\centering
\vspace{-0.1in}
\begin{adjustbox}{width=\linewidth}
  \begin{tabular}{c|l}
\toprule
Original query  & 0-dimensional biomaterials show inductive properties.  \\ 
\midrule
Rephrased query &Do {\color{blue}nano}-sized biomaterials possess unique properties that can trigger specific reactions in biological systems?
\\ 
\midrule
\multirow{8}{*}{Answer document}  & 'title': 'New opportunities: the use of {\color{blue}nano}technologies to manipulate and track stem cells.'\\ 
&'text': '{\color{blue}Nano}technologies are emerging platforms that could be useful in measuring, \\ 
& understanding, and manipulating stem cells. Examples include magnetic {\color{blue}nano}particles and quantum dots \\ 
& for stem cell labeling and in vivo tracking; {\color{blue}nano}particles, carbon {\color{blue}nano}tubes, and polyplexes \\
&for the intracellular delivery of genes/oligonucleotides and protein/peptides; \\ 
&and engineered {\color{blue}nano}meter-scale scaffolds for stem cell differentiation and transplantation. \\ 
&This review examines the use of {\color{blue}nano}technologies for stem cell tracking, differentiation, and transplantation. \\ 
&We further discuss their utility and the potential concerns regarding their cytotoxicity.',          \\ 
\midrule
Original query  & what is the origin of COVID-19 \\ 
\midrule
Rephrased query & What {\color{blue}molecular} {\color{blue}evidence} supports {\color{blue}bats} and pangolins as the likely origin {\color{blue}hosts} of the COVID-19 {\color{blue}virus}?  \\ 
\midrule
\multirow{12}{*}{Answer document}    & 'title': 'Isolation and characterization of a bat SARS-like corona{\color{blue}virus} that uses the ACE2 receptor'\\ 
&'text': 'The 2002–3 pandemic caused by severe acute respiratory syndrome corona{\color{blue}virus} (SARS-CoV) \\ 
&\dots syndrome corona{\color{blue}virus} (MERS-CoV)(2) suggests that this group of {\color{blue}virus}es remains a major threat and that their distribution \\ 
&is wider than previously recognized. Although {\color{blue}bats} have been suggested as the natural reservoirs of both {\color{blue}virus}es(3–5), attempts  \\ 
&to isolate the progenitor {\color{blue}virus} of SARS-CoV from {\color{blue}bats} have been unsuccessful. Diverse SARS-like corona{\color{blue}virus}es (SL-CoVs) \\ 
&have now been reported from {\color{blue}bats} in China, Europe and Africa(5–8), but none are considered a direct progenitor of SARS-CoV \\ 
&because of their phylogenetic disparity from this {\color{blue}virus} and the inability of their spike proteins (S) to use the SARS-CoV \\ 
&cellular receptor {\color{blue}molecule}, the human angiotensin converting enzyme II (ACE2)(9,10). \\ 
&Here, we report whole genome sequences of two novel {\color{blue}bat} CoVs from Chinese horseshoe {\color{blue}bats} (Family: Rhinolophidae)\\  
&in Yunnan, China; RsSHC014 and Rs3367. These {\color{blue}virus}es \dots which has typical corona{\color{blue}virus} morphology, \dots tropism. \\ 
&Our results provide the strongest {\color{blue}evidence} to date that Chinese horseshoe {\color{blue}bats} are natural reservoirs of SARS-CoV, \\ 
&and that intermediate {\color{blue}hosts} may not \dots' \\ 
    \bottomrule
    \end{tabular}
\label{table:2}
\end{adjustbox}
\vspace{-0.15in}
\end{table*}
\begin{table}[t]
\vspace{-0.1in}
\caption{\small\textbf{Ablation study results on SciFact.} This table presents the performance impact of excluding expansion component and optimization component from QOQA, illustrating the importance of each module, in enhancing retrieval accuracy. All scores denote nDCG@10 value.}
\centering
\vspace{-0.1in}
\begin{adjustbox}{width=\linewidth}
  \begin{tabular}{lcc} 
        \toprule
        &QOQA (BM25 score)& QOQA (Dense score) \\\midrule
\multicolumn{3}{l}{\textit{Sparse Retrieval}}\\
Ours         & 67.5 & \textbf{69.7}  \\
w/o expansion    & 65.6   & 66.0    \\
w/o optimization & \textbf{67.6}    & 67.6   \\ \midrule
\multicolumn{3}{l}{\textit{Dense Retrieval}}\\
Ours    & \textbf{75.4}  & \textbf{74.3}\\
w/o expansion    & 72.9  & 74.2 \\
w/o optimization & 73.2  & 72.6 \\ \bottomrule
    \end{tabular}
\label{table:3}
\end{adjustbox}
\vspace{-0.2in}
\end{table}
\paragraph{Baseline}
(1) Sparse Retrieval: (a) BM25~\citep{10.1561/1500000019} model is a widely-used bag-of-words retrieval function that relies on token-matching between two high-dimensional sparse vectors, which use TF-IDF token weights. We used default setting from Pyserini~\citep{Lin_etal_SIGIR2021_Pyserini}. (b) BM25+RM3~\citep{10.1561/1500000019, 10.1145/1645953.1646259} is query expansion method using PRF. We also include (c) BM25+Q2D/PRF~\citep{10.1561/1500000019, jagerman2023queryexpansionpromptinglarge} that use both LLM-based and PRF query expansion methods.
(2) Dense Retrieval: (a) BGE-base-en-v1.5 model is a state-of-the-art embedding model designed for various NLP tasks like retrieval, clustering, and classification. For dense retrieval tasks, we added 'Represent this sentence for searching relevant passages:' as a query prefix, following the default setting from Pyserini.~\citep{Lin_etal_SIGIR2021_Pyserini}. 
We also used CSQE~\citep{lei2024corpus} for both sparse retrieval and dense retrieval.

\paragraph{Implementation details} We utilize GPT-3.5-Turbo~\citep{openai2024gpt4technicalreport} as the LLM optimizer. The temperature is set to 1.0. We set the max optimization iteration as $i = 1, 2, \cdots, 50$. We use $N=5$, $K=3$, $R_0=3$, and $R_i=1$. All hyper-parameters of $k_1=1.2$, $b=0.75$, and $\alpha=0.1$ are set to default values from Pyserini~\citep{Lin_etal_SIGIR2021_Pyserini}.

\paragraph{Retrieval results compared to baselines}
Table~\ref{table:1} illustrates the performance of various document retrieval models across the SciFact, Trec-Covid, and FiQA datasets. For dense retrieval, our enhanced models (+QOQA variants) exhibit superior performance. Notably, QOQA (BM25 score) achieves the best result in SciFact with a score of 75.4, demonstrates strong performance in Trec-Covid with a 79.2 with hybrid score. The consistent performance gain of our QOQA across different datasets highlights effectiveness in improving retrievals.

\paragraph{Case Analysis}
As shown in Table~\ref{table:2}, rephrased queries generated with QOQA are more precise and concrete than the original queries. When searching for the answer document, queries generated with our QOQA method include precise keywords, such as "nano" or "molecular evidence," to retrieve the most relevant documents. This precision in keyword usage ensures that the rephrased queries share more common words with the answer documents. Consequently, the queries utilizing QOQA demonstrate effectiveness in retrieving documents that contain the correct answers, highlighting the superiority of our approach in retrieval tasks.

\paragraph{Ablation Studies}

In our ablation study, we evaluate the impact of the expansion and optimization components in QOQA using both BM25 and Dense scores by systematically removing each component and observing the nDCG@10 results. We remove the document expansion (Blue text in the Figure~\ref{figure:2}) in the "w/o expansion" setup while retaining the optimization step. In the "w/o optimization" setup, we use single-step optimization as $i=1$. As shown in Table~\ref{table:3}, the optimization step improves the search for better rephrased queries. Moreover, without the expansion component, performance significantly drops, especially with the BM25 score. This demonstrates the critical role of the expansion component in creating high-quality rephrased queries and enhancing document retrieval.
\section{Conclusion}
In this paper, we tackled the issue of hallucinations in Retrieval-Augmented Generation (RAG) systems by optimizing query generation. Utilizing a top-k averaged query-document alignment score, we refined queries using Large Language Models (LLMs) to improve precision and computational efficiency in document retrieval. Our experiments demonstrated that these optimizations significantly reduce hallucinations and enhance document retrieval accuracy, achieving an average gain of 1.6\%. This study highlights the significance of precise query generation in enhancing the dependability and effectiveness of RAG systems. Future work will focus on integrating more advanced query refinement techniques and applying our approach to a broader range of RAG applications.

\bibliography{reference}

\end{document}